\documentclass[11pt]{article}

\newcommand{\beq}{\begin{eqnarray}}
\newcommand{\eeq}{\end{eqnarray}}

\newcommand{\be}{\begin{equation}}
\newcommand{\ee}{\end{equation}}
\newcommand{\lwrsim}{\raise0.3ex\hbox{$<$\kern-0.75em\raise-1.1ex\hbox{$\sim$}}}

\def\eq#1{Eq.~(\ref{#1})}

\def\simquand#1{\mbox{\raisebox{-1.2ex}[0.ex][1.6ex]{$\widetilde{\scriptstyle #1}$}}}

\def\goesto#1{\mbox{\raisebox{-1.2ex}[0.ex][-1.4ex]{$\overrightarrow{\scriptstyle #1}$}}}
\def\frsection#1{\section{#1}\hspace*{\parindent}}
\def\sfrac#1#2{\frac{\scriptstyle{ #1}}{\scriptstyle {#2}}}
\usepackage{axodraw}
\usepackage{amsfonts}
\usepackage{amsmath}
\usepackage{amssymb}
\usepackage{latexsym}
\usepackage{color}
\usepackage{colordvi}
\usepackage{epsfig}
\usepackage{array}
\usepackage{pifont}
\usepackage{geometry}
\begin{document}

\title{Divergent  IR gluon  propagator from Ward-Slavnov-Taylor 
identities?}
\author{ Ph.~Boucaud$^a$, J.P.~Leroy$^a$, A.~Le~Yaouanc$^a$, A.Y.~Lokhov$^b$,\\
J. Micheli$^a$, O. P\`ene$^a$, J.~Rodr\'iguez-Quintero$^c$ and 
C.~Roiesnel$^b$ }

\date{}

\maketitle

\begin{center}
$^a$Laboratoire de Physique Th\'eorique et Hautes
Energies\footnote{Unit\'e Mixte de Recherche 8627 du Centre National de 
la Recherche Scientifique}\\
Universit\'e de Paris XI, B\^atiment 211, 91405 Orsay Cedex,
France\\
$^b$ Centre de Physique Th\'eorique\footnote{
Unit\'e Mixte de Recherche 7644 du Centre National de 
la Recherche Scientifique}\\ 
de l'Ecole Polytechnique\\
F91128 Palaiseau cedex, France\\ 
$^c$ Dpto. F\'isica Aplicada, Fac. Ciencias Experimentales,\\
Universidad de Huelva, 21071 Huelva, Spain.
\end{center}

\begin{abstract}

We exploit the Ward-Slavnov-Taylor identity relating the 3-gluons to the
ghost-gluon  vertices to conclude either that the ghost dressing function is  
finite and non vanishing at zero momentum while the gluon propagator diverges 
(although it may do so weakly enough not to be in contradiction with current lattice data)
or that the 3-gluons vertex is non-regular when one momentum goes to zero.
We stress that those results should be kept in mind when one studies the Infrared 
properties of the ghost and gluon propagators, for example by means of 
Dyson-Schwinger equations.

\end{abstract}

{\begin{flushright}
{\small UHU-FP/07-09}\\
{\small CPHT RR 007.0207}\\
{\small LPT-Orsay/07-03}\\
\end{flushright}


\frsection{Introduction}

The infrared behaviour of Landau gauge Green functions is a very active and hot
subject. Dyson-Schwinger (DS) equations have been intensively studied but {\it
the consequences  of the Ward-Slavnov-Taylor identities (WSTI) 
have been largely overlooked}.  It turns out that, after some regularity assumptions are made
on the ghost and gluon vertex functions, {\it they do provide extremely strong
constraints on the zero momentum limit of both the ghost and the gluon
propagator}, namely that the ghost dressing function is finite non vanishing
and the gluon propagator diverges\footnote{This divergence can however be so soft  as  not to contradict the apparent finiteness 
previously stated from lattice data.}.  
The derivation of these results is indeed rather simple and is the main goal of
this letter.

We have already presented in previous  publications~(\cite{Boucaud:2006if,Boucaud:2007va}) arguments in favor of a finite 
non-vanishing ghost dressing function at zero momentum. Notice that this was partly
based  on the study of Dyson-Schwinger equations. Now, although the Dyson Schwinger equations are exact, their practical 
implementation always involves various approximations and hypotheses  which cast a doubt on the general 
validity of the results so obtained. On the contrary, using the WSTI appears to be 
quite simple and to require only a minimum amount of extra information on the regularity of the vertex 
functions. In our opinion this circumstance puts
its consequences on a very firm ground and we think any acceptable solution for the propagators should 
comply with them. As shown in ref.~\cite{Boucaud:2006if}, a non-vanishing ghost dressing function could be 
also favored by the analysis of the Ghost propagator Dyson-Schwinger equation. 
Thus, one is led to conclude either that the gluon propagator diverges and
the ghost dressing is finite non-vanishing or that the regularity hypotheses on the vertices 
should fail.

\frsection{Ward-Slavnov-Taylor identity and the infrared behaviour of the gluon propagator}\label{STID}
The ghost-gluon vertex $\widetilde{\Gamma}_\mu$  is written as
\beq
\widetilde{\Gamma}_\mu^{abc}(p,q;r) &=&  g_0     f^{abc} \ \widetilde{\Gamma}_\mu(p,q;r)
= g_0  (-i p_\nu)  f^{abc} \widetilde{\Gamma}_{\nu\mu}(p,q;r), 
\eeq
where $-p$ ($q$) is the outgoing (incoming) ghost momentum and $r$ is that of 
the gluon, its tensorial structure defined by the following general 
decomposition~\cite{Ball:1980ax}: 
\beq\label{Def2}
 \widetilde{\Gamma}_{\nu\mu}(p,q;r) &=& 
\delta_{\nu\mu} \ a(p,q;r) \ - \ r_\nu q_\mu \ b(p,q;r) 
\ + \ p_\nu r_\mu \ c(p,q;r)\nonumber
\\ 
&+& r_\nu p_\mu \ d(p,q;r) \ + \  p_\nu p_\mu \ e(p,q;r) \ .
\eeq}
The ghost-gluon vertex is related to the  
3-gluon vertex, $\Gamma^{abc}_{\lambda\mu\nu}(p,q,r)$, through the Ward-Slavnov-Taylor identity (\cite{Slavnov,Taylor}): 
\begin{equation}
\label{STid}
\begin{split}
p_\lambda\Gamma_{\lambda \mu \nu} (p, q, r) & =
\frac{F(p^2)}{G(r^2)} (\delta_{\rho\nu} r^2 - r_\rho r_\nu) \widetilde{\Gamma}_{\rho\mu}(r,p;q) 
\\ & -
\frac{F(p^2)}{G(q^2)} (\delta_{\rho\mu} q^2 - q_\rho q_\mu) \widetilde{\Gamma}_{\rho\nu}(q,p;r) \ .
\end{split}
\end{equation}
In this equation $F$ and $G$ are the ghost and gluon dressing functions, defined respectively 
as 
\beq
&&\langle c \bar{c}\rangle = \frac{F(q^2)}{q^2} \ \ \ \ \ \ \mbox{\rm and} \ \nonumber \\ 
&&\langle A_\mu A_\nu \rangle = \frac{1}{q^2} 
\left[ G(q^2)\,(\delta_{\mu\nu} - 
\frac{q_\mu \,q_\nu}{q^2})+(1-\alpha ) \frac{q_\mu \,q_\nu}{q^2} \right]  \ , 
\eeq
with $\alpha$ the usual gauge fixing parameter. We recall in this respect that the WSTI 
holds in any covariant gauge and that the longitudinal part of the propagator is trivial.
We  now make the hypothesis that  $ \Gamma_{\lambda \mu \nu} (p, q, r)$~\footnote{Actually this assumption 
regards only the {\it longitudinal part }, {\it  i.e.} with at least one longitudinal gluon,  of the vertex function (see ref.~\cite{Ball:1980ax} for its 
definition) since the purely transverse one trivially  disappears from the STI.} has a well defined limit 
when anyone  of  its arguments goes to 0, the other ones being kept non-vanishing. Note that this restricts but does not forbid the possible presence 
of singularities in the coefficient functions since they could be compensated by kinematical zeroes stemming 
from the basis tensors. Indeed, this assumption 
and WSTI are all one needs to conclude that the gluon 
propagator diverges at zero momentum. The two transversal projectors in the r.h.s. of \eq{STid} imply that 
a well defined limit at zero momentum for the  l.h.s, after contraction with $r_\nu$ or $q_\mu$, can only 
be zero. For instance, multiplying eq.~(\ref{STid}) with $r_\nu$ gives
\beq\label{STidtral}  
p_\lambda r_\nu \Gamma_{\lambda \mu \nu} (p,q, r)  = -
F(p^2)\frac{q^2}{G(q^2)} (\delta_{\rho\mu}  - \frac{q_\rho q_\mu}{q^2}) r_\nu\widetilde{\Gamma}_{\rho\nu}(q,p;r) \ .
\eeq
Since we suppose that the left hand side has a well defined limit when $q$ goes to zero, the same has 
to be true for the right hand side. 
However, the vector
\beq
r_\nu \widetilde{\Gamma}_{\rho \nu}(q,p;r) \ \equiv \ X(q,p;r) \ r_\rho + Y(q,p;r) \ q_\rho \ ,
\eeq 
where $X$ and $Y$ are combinations of the form factors introduced earlier in \eq{Def2}, 
\beq\label{combi}
X(q,p;r) &=& a(q,p;r)- (r \cdot p) \ b(q,p;r) + (r \cdot q) \ d(q,p;r) \nonumber \\
Y(q,p;r) &=& r^2 c(q,p;r) + (r \cdot q) \ e(q,p;r) \ ,
\eeq
after contraction 
with the tensor $\delta_{\rho\mu}  - \sfrac{q_\rho q_\mu}{q^2}$,
\beq
\left(\delta_{\rho\mu} - \frac{q_\rho q_\mu}{q^2}\right) r_\nu \widetilde{\Gamma}_{\rho \nu}(q,p;r) \ = \ 
\left(r_\mu - \frac{(q \cdot r)}{q^2} q_\mu \right) X(q,p;r)
\eeq
gives an explicit dependence on the direction of $q$. This is in contradiction with a well defined limit 
unless the zero-momentum limit of both sides in \eq{STidtral} is 0.

It is worth noticing that the same conclusion  can be otherwise proven by exploiting the following 
general property of the  3-gluon vertex:
\beq
p_\lambda q_\mu r_\nu \Gamma_{\lambda \mu \nu}(p,q,r) = 0 \ .
\eeq
This last result can be straightforwardly derived from WSTI, and it is  
supported by the perturbative results for the 3-gluon vertex in ref.~\cite{Davydychev:1996pb}.
Then, as $-p=q+r$
\beq
q_\lambda q_\mu r_\nu \Gamma_{\lambda \mu \nu}(p,q,r) + 
r_\lambda q_\mu r_\nu \Gamma_{\lambda \mu \nu}(p,q,r) = 0 \ .
\eeq
Thus, simply  by considering the leading behaviour as $q \to 0$ 
one proves that:
\beq
\lim_{q \to 0} r_\lambda r_\nu \Gamma_{\lambda \mu \nu}(-q-r,q,r) \ = \ 
r_\lambda r_\nu \Gamma_{\lambda \mu \nu}(-r,0,r) = 0 \ .
\eeq
In the usual notation and for a general tensorial decomposition of the 3-gluon vertex 
(see e.g. ref.~\cite{Chetyrkin:2000dq}) this is nothing else 
than the known result $T_3(p^2)=0$.   Equipped with this result (which is valid, of course, 
when any of the arguments goes to 0) and with our previous hypothesis that $\Gamma_{\lambda \mu \nu}$ 
has a well defined limite when anyone of its arguments goes to 0, we can  
conclude that the zero-momentum limit for the l.h.s. of \eq{STidtral} is 
null, {\it i.e.} 
\beq
\lim_{q \to 0} \ p_\lambda r_\nu \Gamma_{\lambda \mu \nu}(p,q,r) \ = - \ 
r_\lambda r_\nu \Gamma_{\lambda \mu \nu}(-r,0,r) \ = \ 0 \ ,
\eeq
and, of course, the same for the r.h.s. of \eq{STidtral},
\beq
\lim_{q \to 0} \left[ F(p^2) \frac{q^2}{G(q^2)} \ 
\left(r_\mu - \frac{(q \cdot r)}{q^2} q_\mu \right) X(q,p;r)
\right] \ = \ 0 \ .
\eeq
This,  in turn, implies that
\beq
\lim_{q \to 0} \frac{q^2}{G(q^2)} \ = \ 0 \ ,
\eeq
{\it i.e.} that the gluon propagator must diverge in the infrared, 
unless $X(q,p;r)$ itself goes to zero.  However this is certainly 
not the case for large enough values of $p^2$ when the perturbative results of Davydychev, 
Osland and Tarasov (\cite{Davydychev:1996pb}) can be used.

For the Ward-Slavnov-Taylor identity to be satisfied there is a compatibility condition  which {\it does not} 
involve the 3-gluon vertex (cf ref.~\cite{Ball:1980ax}). 
Applying the scalar $X$ introduced earlier in eq.~(\ref{combi}) it reads
\begin{equation}\label{Ghexp}
\frac{F(q^2)}{G(p^2)}X(p,q;r)=\frac{F(r^2)}{G(p^2)}X(p,r;q)
\end{equation} 
and has to be satisfied for all $p$'s, which allows to get rid of the $G(p^2)$  denominators.  
Let us consider this relation in the small $r$ limit. The  $X$-factor on the left is the same 
one (except for the interchange of $p$ and $q$)  that appeared in the r.h.s. of \eq{STidtral} 
and remains finite in view of the hypothesis made concerning $\Gamma_{\lambda\mu\nu}$.  This 
implies that $F(r^2) X(p,r;q)$ too remains finite, which implies a strong correlation between the 
infrared behaviours of the ghost propagator and of the ghost-ghost-gluon vertex.

To summarize:

\begin{itemize}
\item {\it  $ \Gamma_{\lambda \mu \nu} (p,q, r)$ infrared-finite and  
$X(q,p;-p-q)\neq 0 \ \Longrightarrow  \ \sfrac{G(q^2)}{q^2} \ \goesto{q\to 0} \ \infty$}
  
\item {\it There exists a strong relationship between  IR behaviours of the ghost propagators and of the ghost-gluon vertex : 
$F(r^2) \propto X(p,q;r)/X(p,r;q)$ when $r \to 0$ and $F(0)$  is presumably finite non-zero}.
\end{itemize}

\frsection{Discussion and conclusion}
  
The Ward-Slavnov-Taylor identity, supplemented with reasonable regularity assumption for the 3-gluon vertex  imposes 
that the gluon propagator is infrared divergent, however slowly this may be\footnote{In the eventuality that not only 
$\Gamma_{\lambda \mu \nu} (p,q, r)$ but the scalar form-factors involved in it would be  finite the divergence could be 
still stronger, including the possibility of a divergent dressing-function.} .  The behaviour of the dressing function 
in this region is usually described using an ``infrared exponent'' :  
\beq
G(q^2) \ \simquand{q^2\to 0} \ (q^2)^{\alpha_G} \ .
\eeq
A divergent propagator would imply either that $\alpha_G$ is smaller than one or that the power law is modified by 
logarithmic corrections. There is no such divergence on the lattice : the propagator at zero momentum is obviously finite and, actually, it has been 
measured directly (our own simulations presently give a value of $\alpha_G$ close to one but do not allow to 
exclude any of the mentioned possibilities). Therefore the divergence could only  manifest itself through the volume dependence. 
The Adelaide group has performed a detailed analysis  of this dependence (cf ref.~ \cite{Bonnet:2001uh}) 
and its results rather favours a finite IR propagator.  
Very recently, some authors~\cite{Silva:2005hb} have pointed that, after analyzing very-low-momentum data 
obtained from large asymmetric lattices, they found a vanishing gluon propagator to be favoured. However, that 
result is not confirmed by the authors of ref.~\cite{Ilgenfritz:2006he,Sternbeck:2005re}. 
We neither agree with that conclusion and discuss about that and their implications in ref.~\cite{Boucaud:2006pc}.
While the WSTI and its consequences hold in all covariant gauges 
those observations have been made in the special case of the Landau gauge where there are of course no longitudinal gluons 
although the WSTI involves  the non transverse vertex function; in that sense it appears as  some kind of  ``limiting case'' of 
the general covariant gauge. One could therefore think of a divergent term whose coefficient would vanish as $1-\alpha$ as 
it is the case for instance for the one-loop anomalous dimension of the ghost-gluon vertex. But this would lead to a hardly acceptable discontinuity in the r.h.s of equation~(\ref{STid})  when $\alpha \to 1$.  A simpler and --maybe--  
more natural explanation would be to imagine that the rate of divergence is too slow to be seen at present.
 
Regarding the ghost infrared exponent,  we have to make an hypothesis about the way    
the combination  of scalars, $X(p,r;q)$ defined by \eq{combi}, in the right hand side of equation~(\ref{Ghexp}) 
behaves at small-$r$.  Davydychev's formulas show that, at one loop, both $b$ and $d$ suffer of a logarithmic divergence 
in this  limit. The one in $b$ would be compensated by the $(r.p)$ factor in front of it, but not the one in $d$.  
In any case we do not know what the situation is for the full non-perturbative quantities. If $X$ goes to some finite non-zero 
limit as $r$ goes to zero $F(r^2)$ will also do.  This is the situation we had considered in  
ref.~\cite{Boucaud:2006if,Boucaud:2007va} and it implies that $\alpha_F$ is zero. 
If, on the contrary, the divergence persists $F$ will have to go to zero with $r$. 
The remarks we have made previously regarding the Landau gauge apply of course also for the ghost propagator but at least 
the finiteness of the dressing function seems to be on a safe ground since one should have to imagine  a divergent 
part with a coefficient proportional to  $\delta(1-\alpha)$. 
 
   Some of the results we have presented in this note are not new (for instance  the conclusion that  F(0) is finite and 
non  zero  can be found in  ref.~\cite{Chetyrkin:2000dq}) but it seems that their consequences have often been overlooked. 
The arguments we have presented rest exclusively  on the Ward-Slavnov-Taylor identity with which,  we think, 
any sensible solution,  among which the ones derived from  Schwinger-Dyson equations, should comply. 
In view of our own experience with lattice simulations, {\bf our preference would go to a weakly divergent 
gluon propagator together with a finite and non-zero ghost dressing function}.  Of course simulations 
with large lattices would be necessary to discriminate unambiguously between the various possibilities, 
as would be the use  of a general covariant gauge to cope with the would-be special characteristics of 
the Landau gauge.

\paragraph{Acknowledgements:} This work has been partially supported by the spanish M.E.C. project 
FPA2006-13825.

\vspace{2cm}
\bibliographystyle{unsrt}\bibliography{SDST}

\begin{thebibliography}{10}

\bibitem{Boucaud:2006if}
Ph. Boucaud et~al.
\newblock Is the {QCD} ghost dressing function finite at zero momentum?
\newblock {\em JHEP}, 06:001, 2006.

\bibitem{Boucaud:2007va}
Ph. Boucaud et~al.
\newblock Constraints on the ir behaviour of gluon and ghost propagator from
  ward-slavnov-taylor identities.
\newblock {\em Eur. Phys. J.}, A:in presse, 2007.

\bibitem{Ball:1980ax}
James~S. Ball and Ting-Wai Chiu.
\newblock Analytic properties of the vertex function in gauge theories. 2.
\newblock {\em Phys. Rev.}, D22:2550, 1980.
\newblock ERRATUM ibid 23(1981),3805.

\bibitem{Slavnov}
A.~A. Slavnov.
\newblock Ward identities in gauge theories.
\newblock {\em Theor. Math. Phys. 10 (1972) 99[Teor. Mat. Fiz. 10 (1972)
  153].}, 1972.

\bibitem{Taylor}
J.~C. Taylor.
\newblock Ward identities and charge renormalization of the {Y}ang-{M}ills
  field.
\newblock {\em Nuclear Physics B33, 436}, 1971.

\bibitem{Davydychev:1996pb}
P.~Osland A.~I.~Davydychev and O.~V. Tarasov.
\newblock Three-gluon vertex in arbitrary gauge and dimension.
\newblock {\em Phys. Rev. D 54, 4087 (1996)}, 1996.
\newblock Erratum ibid. D 59, 109901 (1999).

\bibitem{Chetyrkin:2000dq}
K.~G. Chetyrkin and A.~Retey.
\newblock Three-loop three-linear vertices and four-loop {MOM} {$\beta$}
  functions in massless {QCD}.
\newblock hep-ph/0007088, 2000.

\bibitem{Bonnet:2001uh}
Frederic D.~R. Bonnet, Patrick~O. Bowman, Derek~B. Leinweber, Anthony~G.
  Williams, and James~M. Zanotti.
\newblock Infinite volume and continuum limits of the {L}andau-gauge gluon
  propagator.
\newblock {\em Phys. Rev.}, D64:034501, 2001.

\bibitem{Silva:2005hb}
P.~J. Silva and O.~Oliveira.
\newblock Infrared gluon propagator from lattice qcd: results from large
  asymmetric lattices.
\newblock {\em Phys. Rev.}, D74:034513, 2006.

\bibitem{Ilgenfritz:2006he}
E.~M. Ilgenfritz, M.~Muller-Preussker, A.~Sternbeck, A.~Schiller, and I.~L.
  Bogolubsky.
\newblock Landau gauge gluon and ghost propagators from lattice qcd.
\newblock {\em hep-lat/0609043 (IRQCD06)}, 2006.

\bibitem{Sternbeck:2005re}
A.~Sternbeck, E.~M. Ilgenfritz, M.~Muller-Preussker, and A.~Schiller.
\newblock Landau gauge ghost and gluon propagators and the faddeev- popov
  operator spectrum.
\newblock {\em Nucl. Phys. Proc. Suppl.}, 153:185--190, 2006.

\bibitem{Boucaud:2006pc}
Ph. Boucaud et~al.
\newblock Short comment about the lattice gluon propagator at vanishing
  momentum.
\newblock {\em hep-lat/0602006}, 2006.

\end{thebibliography}

\addcontentsline{toc}{section}{References}
\end{document}